\documentclass[a4paper,11pt]{article}
\pdfoutput=1
\usepackage{jheppub}
\usepackage{amssymb}
\usepackage{amsmath}
\usepackage{mathtools}
\usepackage{amsfonts}
\usepackage{dsfont}
\usepackage{young}
\usepackage[vcentermath]{youngtab}
\usepackage{bm}
\usepackage{braket}
\usepackage{simplewick}
\usepackage[offset=1.25em]{simpler-wick}
\usepackage{bm}
\usepackage[most]{tcolorbox}
\usepackage{enumerate}
\usepackage{xcolor}

\tcbset{colback=yellow!10!white, colframe=red!50!black,
        highlight math style= {enhanced, 
            colframe=red,colback=red!10!white,boxsep=0pt}
        }

\usepackage{mathrsfs}
\usepackage{tikz}
\usetikzlibrary{shapes}
\usetikzlibrary{arrows.meta}
\usetikzlibrary{positioning}
\usetikzlibrary{positioning}
\usetikzlibrary{decorations.markings}
\usetikzlibrary{decorations.pathmorphing}

\usepackage{comment}

\numberwithin{equation}{section}

\DeclareMathAlphabet{\mathpzc}{OT1}{pzc}{m}{it}

\newcommand{\mrm}[1]{\mathrm{#1}}
\newcommand{\mcl}[1]{\mathcal{#1}}
\newcommand{\mbb}[1]{\mathbb{#1}}

\newcommand{\mfr}[1]{\mathfrak{#1}}

\newcommand{\cint}[1]{\underset{#1}{\oint}}
\newcommand{\llangle}{\left\langle}
\newcommand{\rrangle}{\right\rangle}
\newcommand{\bbr}[1]{\big(#1\big)}

\newcommand{\be}{\begin{equation}}
\newcommand{\ee}{\end{equation}}

\definecolor{bcolor}{HTML}{135424}

\title{Anomalous dimensions in the symmetric orbifold}
\author[a]{Matthias R.~Gaberdiel,}
\author[b]{Felix Lichtner,}
\author[a]{and Beat Nairz}
\affiliation[a]{Institut f\"ur Theoretische Physik,
ETH Z\"urich,\\
Wolfgang-Pauli-Strasse 27,
8093 Z\"urich, Switzerland}
\affiliation[b]{Max Planck Institute for Gravitational Physics (AEI) \\
Am M\"uhlenberg 1, D-14476 Potsdam, Germany}
\emailAdd{gaberdiel@itp.phys.ethz.ch}
\emailAdd{felix.lichtner@aei.mpg.de}
\emailAdd{nairzb@ethz.ch}

\abstract{Recently, the anomalous conformal dimensions of the symmetric orbifold under the $2$-cycle twisted sector deformation were calculated using the perturbed action of the supercharges. In particular, explicit and simple formulae for the dispersion relations of the torus magnons in the $w$-cycle twisted sector were derived for large $w$. In this paper we reproduce these results from a direct perturbed $2$-point function calculation. In the process we also develop techniques (and a Mathematica code) that allows one to do these calculations for arbitrary quarter BPS states at finite $w$.}

\begin{document}
\maketitle

\section{Introduction}

The symmetric orbifold of $\mathbb{T}^4$ is dual to string theory on ${\rm AdS}_3\times {\rm S}^3 \times \mathbb{T}^4$ with one unit of NS-NS flux \cite{Gaberdiel:2018rqv,Eberhardt:2018ouy,Eberhardt:2019ywk}. Recently, the perturbation that corresponds to switching on R-R flux in the dual AdS background was studied systematically in the symmetric orbifold \cite{Gaberdiel:2023lco}, and the expected integrable structure of  \cite{Babichenko:2009dk,Hoare:2013lja,Borsato:2013qpa,Lloyd:2014bsa,Frolov:2023pjw} was recovered.

The analysis in \cite{Gaberdiel:2023lco} was performed by determining the action of the supercharges on the various states to first order in perturbation theory, and then inferring the anomalous conformal dimensions from them. This approach generalised and refined previous calculations, see in particular \cite{Gava:2002xb},\footnote{For other attempts to uncover the integrable structure underlying the symmetric orbifold see \cite{Lunin:2002fw,Gomis:2002qi,David:2008yk,Pakman:2009zz,Pakman:2009mi}.} and the structure that was found is very similar to the dynamical spin chain of ${\cal N}=4$ SYM \cite{Beisert:2005tm}, see \cite{Beisert:2010jr} for a review. While this indirect method was very successful for ${\cal N}=4$ SYM, it is quite delicate in the present context since in order for the integrable structure to emerge, it was not only necessary to consider the planar $N\rightarrow\infty$ limit, but also a large charge limit $w\rightarrow \infty$. The latter is somewhat subtle since there are subleading terms  (in $\frac{1}{w}$) in the action of the supercharges on the magnon states. These were systematically neglected in the analysis in \cite{Gaberdiel:2023lco}, but they could, in principle, modify the answer for the anomalous dimensions.

In this paper we shall demonstrate that the approximation scheme of \cite{Gaberdiel:2023lco} is indeed justified. More specifically, we shall determine the anomalous conformal dimensions by a direct second order perturbation theory calculation, and we find that this calculation reproduces nicely the predictions of \cite{Gaberdiel:2023lco}. We should stress that while the second order perturbation theory does not make any approximations, it is much more involved, and can only be performed for relatively small values of $w$.\footnote{Perturbative calculations of this kind had been studied before, see for example \cite{Keller:2019suk,Benjamin:2021zkn,Lima:2020boh,Burrington:2012yq,Guo:2022ifr,Hughes:2023fot}.} On the other hand, the approach of  \cite{Gaberdiel:2023lco} defines a very efficient approximation scheme that allows one to read off the underlying integrable structure, something that would have been very difficult to extract from the full perturbation calculation.

\subsection{The approach of \cite{Gaberdiel:2023lco}}\label{sec:old}

Let us be a bit more specific, and review the basic idea of the calculation in \cite{Gaberdiel:2023lco}. To be concrete and compare to \cite{Gaberdiel:2023lco} as directly as possible, we shall mainly concentrate on the $2$-magnon states of the form
\begin{equation}\label{2magnon}
|\Psi_m\rangle \equiv \tfrac{1}{\sqrt{m(w-m)}}\alpha^2_{-\frac{m}{w}}\alpha^2_{-1+\frac{m}{w}}\ket{w}\ .
\end{equation}
We have also tested our results on other multi-magnon excitations, using the notebook provided.\footnote{States of the form $\alpha^1 \alpha^1\ket{w}$ are related to the ones in eq.~(\ref{2magnon}) by an outer $\mfr{su}(2)$ symmetry, see eq.~(\ref{eq:res_charge}). On the other hand, the anomalous dimensions of states like $\psi^-\psi^-\ket{w}$ are not related by symmetry.} Here, the $\alpha^2$ are free torus bosons; our conventions are collected in Appendix \ref{app:conventions}.
In the analysis of \cite{Gaberdiel:2023lco} the action of a right-moving supercharge $\tilde{S}_2 \equiv \tilde{G}^{'+}_{-1/2}$ on this state was determined
to first order in perturbation theory
\begin{equation}\label{1.2}
\tilde{S}_2\cdot\alpha^2_{-\frac{m}{w}}\alpha^2_{-1+\frac{m}{w}}\ket{w} =
\sum_{n}
c_m^n\,\alpha^2_{-\frac{n}{w+1}}\psi^-_{-\frac{1}{2}+\frac{n}{w+1}}\ket{w+1} +
\sum_{\chi}c_m^\chi \,\ket{\chi}\ ,
\end{equation}
where $\ket{\chi}$ schematically accounts for the states that are not included in the first sum. One can check that the coefficients $c_m^\chi$ are suppressed at least as $\mcl{O}\big(\frac{1}{w}\big)$ for large $w$, see the end of Section~\ref{sec:intermediate} below for more details.

This fact motivated the limit prescription
\begin{equation}\label{eq:inf_w_limit}
\lim_{w\to\infty}\,\tilde{S}_2\cdot\alpha^2_{-\frac{m}{w}}\alpha^2_{-1+\frac{m}{w}}\ket{w}
 \overset{\text{def.}}{=} \lim_{w\to\infty}\,\sum_{n}
c_m^n\,\alpha^2_{-\frac{n}{w+1}}\psi^-_{-\frac{1}{2}+\frac{n}{w+1}}\ket{w+1}\ .
\end{equation}
Moreover, the $c_m^n$ then have a simple and suggestive form which can be
formally continued to $w=\infty$, thereby making contact  with the integrability
description of, for example, \cite{Lloyd:2014bsa}.

By considering the anti-commutator with the conjugate charge
$\tilde{Q}_2 \equiv \tilde{G}^{'-}_{+1/2}$ (for which a similar formula could be derived), the anomalous conformal dimension of (\ref{2magnon}) to second order in perturbation theory was then determined to be
\begin{equation}\label{eq:int_anomalous_dimension}
\tilde{\gamma}(|\Psi_m\rangle) = g^2\,
\frac{\sin^2\!\big(\pi\frac{m}{w}\big)}{\frac{m}{w}\big(1-\frac{m}{w}\big)}\ .
\end{equation}

\subsection{The direct calculation}\label{sec:direct}

It is instructive to compare the above results to a direct determination of the anomalous dimension from calculating the perturbed $2$-point function. The matrix element of $\tilde{\gamma}$ mixing $\ket{\Psi_m}$ and $\ket{\Psi_n}$ can be calculated by integrating the four-point function
\begin{equation}
\bra{\Psi_n}\tilde{\gamma}\ket{\Psi_m} = \left. -\pi g^2\,\int\!
d^2x \,d^2y\,\llangle\Psi_n(z_1,\bar{z}_1)\,\Phi(x,\bar{x})\,\Phi(y,\bar{y})\,\Psi_m(z_2,\bar{z}_2)\rrangle\right|_{{\rm log}|z_1-z_2|^2} \ ,
\end{equation}
where $\Phi$ is the exactly marginal perturbing field,
\begin{equation}\label{eq:perturbing_field}
\Phi = \frac{i}{\sqrt{2}}\,\left(G^-_{-\frac{1}{2}}\tilde{G}'^-_{-\frac{1}{2}} -G'^-_{-\frac{1}{2}}\tilde{G}^-_{-\frac{1}{2}} \right)\sigma_2\ ,
\end{equation}
and $\sigma_2$ is the dimension $h=\tilde{h}=\frac{1}{2}$ BPS state (with spin up) in the 2-cycle twisted sector.

As indicated above, it is only the term proportional to $\log|z_1 - z_2|^2$ that modifies the conformal dimension of $|\Psi_m\rangle$, and it can be extracted as explained in \cite{Keller:2019suk}. This calculation can be done directly using a suitable covering map, and we explain how this works in detail in Section~\ref{sec:4point} below.

As was explained in \cite{Benjamin:2021zkn}, the anomalous dimension matrix can be obtained from the different $3$-point functions that appear in the analysis of eq.~(\ref{1.2}) via
\begin{equation}\label{fullmixing}
\bra{\Psi_n}\tilde{\gamma}\ket{\Psi_m} = \sum_\chi
\bra{\Psi_n}\tilde{Q}_2\ket{\chi}\bra{\chi}\tilde{S}_2\ket{\Psi_m} + \sum_{\chi}
\bra{\Psi_n}\tilde{S}_2\ket{\chi}\bra{\chi}\tilde{Q}_2\ket{\Psi_m}\ ,
\end{equation}
where in each term the sum over $\chi$ accounts for all intermediate states of eq.~(\ref{1.2}), including the terms that are individually suppressed as ${\cal O}(\frac{1}{w})$. In the approximation of \cite{Gaberdiel:2023lco}, all the subleading $\ket{\chi}$ are neglected; then only the two-magnon states $\ket{\Psi_m}$, $m=1,\dots,\lceil {\frac{w-1}{2}}\rceil$ mix among themselves, and the anomalous dimensions are determined by diagonalising the submatrix $\bbr{\bra{\Psi_n}\tilde{\gamma}\ket{\Psi_m}}_{n m}$ of the full mixing matrix $\tilde{\gamma}$.

If we calculate $\tilde{\gamma}$ using the four-point function, we automatically allow for arbitrary intermediate states. While the contribution of each state in the second sum of eq.~(\ref{1.2}) is individually suppressed in $\frac{1}{w}$, there are many such states, and it is not clear whether neglecting them, as in  eq.~(\ref{eq:inf_w_limit}), will still lead to the correct result for the anomalous dimensions. In fact, as we shall see in Section~\ref{sec:2magnon} below, the individual matrix elements are significantly affected by this approximation. By comparing the direct calculation of the left-hand-side with the different $3$-point functions on the right-hand-side, we can explicitly determine which intermediate states contribute even at large $w$, and we identify the additional intermediate states (beyond those that are accounted for by the first sum of eq.~(\ref{1.2})) in Section~\ref{sec:intermediate}. (The relevant states at large $w$ turn out to be certain $4$-magnon states in the $w$-cycle twisted sector.) In view of these additional intermediate contributions, it is then inconsistent to  restrict the external states just to be the $2$-magnon states of the form $|\Psi_m\rangle$. Instead, if we want to determine the anomalous conformal dimensions from the $4$-point mixing matrix, we need to diagonalise a much bigger matrix, that includes at least also these $4$-magnon states as external states. Once this is done, the anomalous conformal dimensions that are determined from diagonalising the full mixing matrix $\tilde{\gamma}$ reproduce nicely the predictions of eq.~(\ref{eq:int_anomalous_dimension}), see Section~\ref{sec:full_AD} below. This is the main result of our paper.


The relevant eigenstates of the anomalous mixing matrix involve in addition to the $2$-magnon `head' of the form (\ref{2magnon}), a `tail' of $4$- (and higher) magnon states. We also evaluate the supercharges on these eigenstates, and we find that, when restricted to the $2$-magnon components, the action reduces to that of \cite{Gaberdiel:2023lco}, see Section \ref{sec:3.3}. Our analysis therefore demonstrates that the large $w$ approximation of  \cite{Gaberdiel:2023lco} captures the relevant algebraic structure (including the anomalous conformal dimensions) of the problem. At the same time, it is very efficient, and unlike the full calculation, see Section~\ref{sec:4point}, one can extrapolate the answer to large $w$.

We should mention that the full mixing matrix calculation is very complicated, and in order to do it efficiently, we had to write a Mathematica code that allows one to calculate the mixing matrix of in principle any two $\frac{1}{4}$-BPS states to second order in perturbation theory. This code may also prove useful in other contexts, and we have therefore included it as an ancillary file to the {\tt arXiv} submission.

\medskip

The paper is organised as follows. In Section~\ref{sec:4point} we explain the `direct calculation' of Section~\ref{sec:direct} in detail and show that the eigenvalues of the full problem reproduce the results of \cite{Gaberdiel:2023lco}. The mixing matrix entries are reobtained in Section~\ref{sec:3point} from sums over intermediate states, i.e.\ as in eq.~(\ref{fullmixing}); in particular, this allows us to identify which intermediate states contribute to the mixing matrix for large $w$. We also comment on the action of the supercharges on the eigenstates there, see Section~\ref{sec:3.3}. Our results are summarised in Section~\ref{sec:concl} and we give an outlook on future directions. There are two appendices to which some of the technical material has been relegated.

\section{The four-point function calculation}\label{sec:4point}

In this section we outline the method of calculating the anomalous dimension using the direct method of Section~\ref{sec:direct}.

\subsection{The method of calculation}\label{sec:4pointcalc}

To order $g^2$, the entries of the anomalous dimension matrix $\tilde{\gamma}$ are given by \cite{Keller:2019suk}
\begin{equation}
\bra{\psi'}\tilde{\gamma}\ket{\psi} = -\pi g^2\,\int\!
d^2x\,\llangle\psi'(\infty)\,\Phi(1)\,\Phi(x,\bar{x})\,\psi(0)\rrangle\ ,
\end{equation}
where $\Phi$ is the perturbation in eq.~(\ref{eq:perturbing_field}), and our conventions are spelled out in Appendix~\ref{app:conventions}. This integral is regulated by cutting out disks around the insertions points. Using
that the states we look at are $\frac{1}{4}$-BPS, this can be further
simplified by turning the surface integral into a contour integral \cite{Keller:2019suk}:
\begin{equation}\label{eq:base_space_AD}
\bra{\psi'}\tilde{\gamma}\ket{\psi} = 2\pi^2
g^2\,\cint{C(0,1,\infty)}\!\!\!\!\!dx\,\llangle\psi'(\infty)\,
\big(\Phi^\dagger_{\mrm{L}}(1)\,\Phi_\mrm{L}(x,\bar{x}) +
\Phi'^\dagger_\mrm{L}(1)\,\Phi'_{\mrm{L}}(x,\bar{x})\big)\,\psi(0)\rrangle\ .
\end{equation}
Here, there are contours around the operator insertions at $0,1,\infty$, and
the operators in the correlator are BPS on the right,
\begin{equation}
\Phi_{\mrm{L}} = G^-_{-\frac{1}{2}}\sigma_2\ ,\quad \Phi'_{\mrm{L}} =
G'^-_{-\frac{1}{2}}\sigma_2\ .
\end{equation}
Furthermore, the dagger acts by charge conjugation, i.e.
\begin{equation}
\Phi_{\mrm{L}}^\dagger = G^+_{-\frac{1}{2}}\sigma_2^\dagger\ ,\quad \Phi'^\dagger_{\mrm{L}} =
G'^+_{-\frac{1}{2}}\sigma_2^\dagger\ ,
\end{equation}
where $\sigma_2^\dagger$ denotes the $h=\tilde{h} = 1/2$ BPS state with spin down, i.e.~$K^3_0\sigma_2^\dagger = \tilde{K}^3_0\sigma_2^\dagger=-\frac{1}{2}\sigma_2^\dagger$.

In the integral above, the contour around $1$ can never contribute, by marginality of $\Phi$ as is also explained in \cite{Keller:2019suk} --- this
will be a consistency check on our calculations later.

\subsubsection{The covering map}
We will be interested in the anomalous dimension matrix, mixing states in the
$w$-twisted sector. The easiest way to calculate the relevant correlators is by using the covering map method \cite{Lunin:2000yv}. For the situation at hand, the ramification profile is $(w,2,2,w)$, and we only consider sphere coverings (since we are interested in the planar large $N$ limit). The covering map can then be described as
\begin{equation}\label{eq:4pt_cov_map}
\bm{\Gamma}(z) = z^w\,\frac{\big(w l -(w-1)\big)z - \big((w+1) l -
w\big)}{z-l}\ .
\end{equation}
Here, we choose to parametrise the covering map by the position of the pole,
$l$, as was also done in \cite{Dei:2019iym} for the special case $w=2$. This has the
advantage that the correlator will be single-valued in $l$, making contour
integration in principle straight-forward.

The covering map has the following properties. It is ramified at
$0,u(l),1,\infty$, with ramification indices $w,2,2,w$, respectively. The
ramification point
unfixed by M\"obius symmetry is
\begin{equation}
u(l) = l\frac{(w+1)l-w}{wl-(w-1)}\ .
\end{equation}
It maps onto the location of the perturbing field
$\Phi\big(x(l),\bar{x}(\bar{l})\big)$
in the base space at
\begin{equation}
x(l)=\bm{\Gamma}\big(u(l)\big) = l^{w-1}
\frac{((w+1)l-w)^{w+1}}{(wl-(w-1))^{w-1}}\ .
\end{equation}
There are also quarter BPS states in the $w+2$ sector which can mix with our `in' states. In order to calculate their matrix elements we thus also need the covering map with ramification profile $(w+2,2,2,w)$; this is given in Appendix \ref{app:wp2_covering_map}.

\subsubsection{The anomalous dimension expressed through the covering map}\label{sec:covering_surface_calculation}

Using the covering map, the calculation of the anomalous dimension in
eq.~(\ref{eq:base_space_AD}) becomes\footnote{In the $x$-contour integral, one needs to sum over all `diagrams' contributing to the correlator, see \cite{Pakman:2009zz}. Carrying out this sum corresponds to integrating over all of $l$-space.}
\begin{equation}
\bra{\psi'}\tilde{\gamma}\ket{\psi} = 2\pi^2 g^2\!\oint
dl\,\frac{dx}{dl}\,\llangle\hat{\psi}'(\infty)\,
\big(\hat{\Phi}^\dagger_{\mrm{L}}(1)\,\hat{\Phi}_\mrm{L}(u,\bar{u})
+
\hat{\Phi}'^\dagger_\mrm{L}(1)\,\hat{\Phi}'_{\mrm{L}}(u,\bar{u})
\big)\,\hat{\psi}(0)\rrangle\ ,
\end{equation}
where the hat denotes the lift of the field to the covering surface. The
contour of integration is around all of the $l$,
s.t.~$x(l)\in\{0,1,\infty\}$, which will be specified below.

The right-moving part of the correlator is the four-point function of
(anti-)chiral primaries\footnote{We have made use of the freedom to shift the
right-moving correlator by a constant to set the regular value at $\infty$ to
zero for convenience.}
\begin{equation}
\llangle \hat{\phi}_w^\dagger(\infty)\,\hat{\phi}_2^\dagger(1)\,
\hat{\phi}_2(\bar{u})\,\hat{\phi}_w(0)\rrangle = \frac{w-1-2w\bar{l}}{2 w (w+1)
(\bar{l}-1)
\big((w+1)\bar{l}-(w-1)\big)}\ ,
\end{equation}
where $\hat{\phi}_w$ denotes the lift of the BPS state $\ket{w}$.

The main computational challenge, however, comes from the left-moving part of the correlator. Consider for
example $\ket{\psi} = \alpha^2_{-\frac{m}{w}}\alpha^2_{-1+\frac{m}{w}}\ket{w}$.
The correlator can then be expressed as
\begin{equation}
\llangle \cdots \hat{\psi}(0)\rrangle =
\cint{C(0)}dz_1\,\bm{\Gamma}^{-\frac{m}{w}}(z_1)\cint{C(0)}dz_2\,
\bm{\Gamma}^{-1+\frac{m}{w}}(z_2)\,\llangle \cdots
\alpha^2(z_1)\alpha^2(z_2)\,\hat{\phi}_w(0)\rrangle\ .
\end{equation}
The right-hand-side can now be computed using the Wick contractions of the
free CFT on the covering surface. Once the $z_i$ contour integrals have been
carried out, one is left with
\begin{equation}
\bra{\psi'}\tilde{\gamma}\ket{\psi} = 2\pi^2 g^2\!\oint
dl\,\frac{dx}{dl}\,F\big(\phi',\phi,l,\bar{l}\big)\ ,
\end{equation}
where $F\big(\phi',\phi,l,\bar{l}\big)$ denotes the resulting correlator. We explain the details of one such sample calculation in Appendix~\ref{sec:explicit_correlator}.

Let us close this section by describing the $l$-contour integral in more detail. The relevant
pre-images of $x=0,\infty$ are
\begin{align}
l_0^- &= 0\ , & l_0^+ &= \frac{w}{w+1}\ ,\nonumber\\
l_\infty^- &= \frac{w-1}{w}\ , & l_\infty^+ &= \infty\ .
\end{align}
These are the only pre-images of $0,\infty$ where the correlator can develop a
singularity, as they correspond to $u(l)=0,\infty$. The superscript denotes the `channel' of the intermediate fusion.
The `$+$' stands for the case where the field which propagates in the middle is in the
$w+1$-twisted sector, the `$-$' stands for the case where it is in the
$w-1$-sector.

Additionally there are two pre-images of $x=1$ where the correlator can have a
pole (corresponding to the diagrams where the twist-2 fields contract, $u(l)=1$).
They are
\begin{equation}
l_1^-=1\ ,\quad l_1^+ = \frac{w-1}{w+1}\ .
\end{equation}
As mentioned before, however, there should not be any contribution from these `self-contractions' of $\Phi$ since the perturbation is exactly marginal. Confirming that this is indeed the case provides a non-trivial consistency check on our calculations.

\subsection{Two-magnon external states}\label{sec:2magnon}

Armed with this formalism, one can then calculate for example the mixing of all
the prototypical states, i.e.~the anomalous dimension submatrix on the space
\begin{equation}\label{V.2}
\mbb{V}_2 := \big\{\alpha^2_{-\frac{m}{w}}\alpha^2_{-1+\frac{m}{w}}\ket{w}\,
\big| \,m=1,\dots,w-1\big\}\ ,
\end{equation}
where the subscript $2$ refers to the fact that we are looking at 2-magnon excitations.

If we diagonalise this matrix, the mixing between different states is small and the eigenvalues are close to the diagonal elements. One observes that these do not approach the predicted spectrum eq.~(\ref{eq:int_anomalous_dimension}) for $w\to\infty$; an essential difference is that the anomalous dimension for $m\to 0$ does not approach $0$. In Fig.~\ref{fig:w16_spectrum}, the diagonal elements for $w=16$ are shown.

As explained above, the reason for this mismatch is that, if we include arbitrary intermediate states (as is automatic in this $4$-point function calculation), we also need to consider the mixing matrix between arbitrary external states.

\subsection{The full anomalous dimension matrix}\label{sec:full_AD}

In view of the above, we therefore have to consider arbitrary matrix elements, i.e.\ we cannot just evaluate the mixing matrix on the space $\mbb{V}_2$ from above. Instead we need to evaluate the mixing matrix on the space
\begin{equation}
\mbb{V}_2\oplus\mbb{V}_4\oplus\mbb{V}_6\oplus\cdots\ ,
\end{equation}
where $\mbb{V}_k$ contains the states with $k$ magnon excitations. Additionally, there are states in the $w+2$-twisted sector which can mix with $\mbb{V}_2$ as well, see Appendix \ref{app:wp2_covering_map}. Given that the perturbation preserves several charges, see Appendix \ref{app:conserved_charges}, only certain families of states can mix with the states in eq.~(\ref{V.2}), and we include in $\mbb{V}_k$ only those states for which this is possible. For example, the relevant four-magnon excitations on top of the BPS vacuum are
\begin{align}
\mbb{V}_4 = \big\{  &\alpha^2_{-\frac{m_1}{w}}\alpha^2_{-\frac{m_2}{w}}\alpha^2_{-\frac{m_3}{w}}\bar{\alpha}^1_{-\frac{m_4}{w}}\ket{w}\ ,\nonumber\\
&\alpha^2_{-\frac{m_1}{w}}\alpha^2_{-\frac{m_2}{w}}\alpha^1_{-\frac{m_3}{w}}\bar{\alpha}^2_{-\frac{m_4}{w}}\ket{w}\ ,\nonumber\\
&\alpha^2_{-\frac{m_1}{w}}\alpha^2_{-\frac{m_2}{w}}\psi^+_{-\frac{1}{2}-\frac{m_3}{w}}\bar{\psi}^-_{\frac{1}{2}-\frac{m_4}{w}}\ket{w}\ ,\nonumber\\
&\alpha^2_{-\frac{m_1}{w}}\alpha^2_{-\frac{m_2}{w}}\bar{\psi}^+_{-\frac{1}{2}-\frac{m_3}{w}}\psi^-_{\frac{1}{2}-\frac{m_4}{w}}\ket{w}\ ,\nonumber\\
&\alpha^2_{-\frac{m_1}{w}}\bar{\alpha}^2_{-\frac{m_2}{w}}\psi^+_{-\frac{1}{2}-\frac{m_3}{w}}\psi^-_{\frac{1}{2}-\frac{m_4}{w}}\ket{w}\ \big|\,m_1+m_2+m_3+m_4 = w\big\}\ .
\end{align}
Because there is more freedom in how the $m_i$ are distributed over the component fields, one sees that the dimension of $\mbb{V}_4$ scales as $\mrm{dim}\,\mbb{V}_4= \mcl{O}\big(w^3\big)$, while $\mrm{dim}\,\mbb{V}_2 = \mcl{O}\big(w\big)$, making the problem less tractable for large $w$. In general, the dimension of $\mbb{V}_k$ scales as $\mcl{O}\big(w^{k-1}\big)$, because there are $k$ undetermined mode numbers and one constraint. However, at any finite $w$, only the spaces up to and including $\mbb{V}_{w+2}$ are non-trivial. So while it is in principle possible to find the full mixing matrix, it quickly becomes unwieldy.

Calculating the anomalous dimension matrix when including these new states proceeds along the same lines as for the example in Appendix \ref{sec:explicit_correlator}. The details are more complicated, however: since there are also fermionic descendants, one needs additional types of contractions with many special cases, as the interaction with the spin-fields makes the fermionic correlators trickier than the bosonic ones. The full calculation thus quickly becomes overwhelming when written down by hand, and we have automated the process in Mathematica, allowing us to calculate any correlator of arbitrary external states. We have included the relevant Mathematica notebook as an ancillary file to the {\tt arXiv} submission.

Since we are interested in relating our results to those of \cite{Gaberdiel:2023lco} we will concentrate in the following on the eigenvalues associated to the two-magnon states $\ket{\Psi_m}$ in eq.~(\ref{2magnon}). We select these by concentrating on the eigenstate $\ket{\epsilon_m}$ of the full problem with maximal overlap $|\langle\Psi_m|\epsilon_m\rangle|$, and we associate its eigenvalue $\epsilon_m$ to $\ket{\Psi_m}$. We have carried out this calculation (including all external states) for small $w\leq 6$ and extracted the spectrum this way. The states $\ket{\epsilon_m}$ turn out to be low-lying in the spectrum. Furthermore, we find that one can obtain a good approximation to the eigenvalues by firstly restricting the mixing within the $w$-twisted sector, and secondly by only including states with two or four magnon excitations, i.e.\ by projecting onto the space $\mbb{V}_2\oplus\mbb{V}_4$. This is sensible, as the mixing between $\mbb{V}_2$ and $\mbb{V}_k$ is suppressed by $w^{1-k/2}$, as we will discuss in Section \ref{sec:intermediate}.

\begin{figure}[ht]
\centering
\includegraphics[scale=0.85]{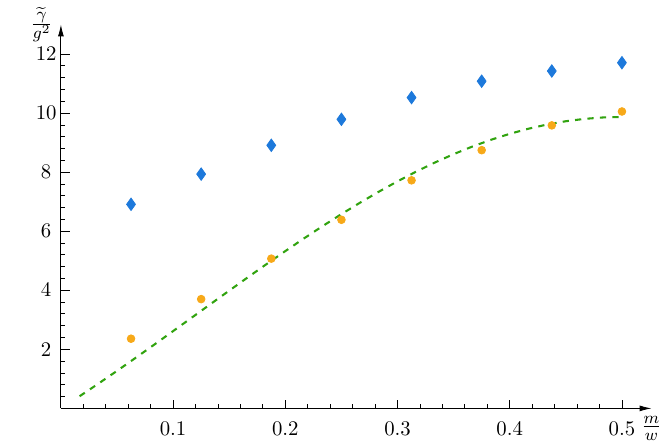}
\caption{The anomalous dimension associated to the two-magnon states as a function of $m$ for $w=16$. Shown are the diagonal elements $\bra{\Psi_m}\tilde{\gamma}\ket{\Psi_m}$ (blue diamonds), the eigenvalues of the full eigenstates with the largest overlap with the respective two-magnon state $\ket{\Psi_m}$ (orange dots), and the (scaled) predicted spectrum for large $w$, see eq.~(\ref{eq:int_anomalous_dimension}) (dashed green line).}
\label{fig:w16_spectrum}
\end{figure}

Using this approximation we can carry out the calculation for reasonably large $w$, although much smaller than in \cite{Gaberdiel:2023lco}. In Fig.~\ref{fig:w16_spectrum}, the result for $w=16$ is shown. One can see that at this intermediate value for $w$, the spectrum is already very well described by the integrability dispersion relation in eq.~(\ref{eq:int_anomalous_dimension}). Explicitly, we find
\begin{equation}\label{eq:approximate_dispersion_relation}
\epsilon_m \approx \frac{g^2}{C}\,\frac{\sin^2\big(\pi \frac{m}{w}\big)}{\frac{m}{w}\big(1-\frac{m}{w}\big)}\ ,
\end{equation}
where $C\approx 0.41$ is some overall factor which can be absorbed by a rescaling of $g$. It is $w$ independent, but we do not have an interpretation for this constant.

Let us stress how much more computationally expensive the full calculation is (relative to the approximation of \cite{Gaberdiel:2023lco}): to obtain the 8 points in Fig.~\ref{fig:w16_spectrum}, roughly 1500 states need to be taken into account. Each entry in the anomalous dimension matrix is computed by taking multiple analytic\footnote{In the computation, rational functions of high degree with large integer coefficients appear. Numerical computation becomes extremely unreliable, therefore the residues need to be taken analytically.} residues, resulting in a large computational effort just to find the result for $w=16$.

\section{The three-point function calculation}\label{sec:3point}

In this section we explain how the calculation of the anomalous dimension matrix using four-point functions can be reproduced from three-point functions as in eq.~(\ref{fullmixing}). The reason to carry out this programme is three-fold. First of all, it allows us to determine which intermediate states effectively (i.e.\ in the large $w$ limit) contribute to the mixing matrix, see Section~\ref{sec:intermediate}, and this then also informs what external states one should consider, see the discussion in Section~\ref{sec:full_AD} above. Secondly, it provides a highly non-trivial consistency check on the different calculations, see in particular Section~\ref{sec:cons}. Thirdly, this algebraic calculation is directly comparable to \cite{Gaberdiel:2023lco}, and we can gain a more microscopic understanding of why the analysis in Section \ref{sec:old} gives the correct result, see Section~\ref{sec:3.3}.

\subsection{The method of calculation}

As mentioned before, the anomalous dimension matrix on quarter BPS states can
be decomposed into three-point functions where the supercharges act on the
states \cite{Gava:2002xb,Benjamin:2021zkn}
\begin{equation}
\bra{\psi'}\tilde{\gamma}\ket{\psi} = \sum_\chi
\bra{\psi'}\tilde{Q}_2\ket{\chi}\bra{\chi}\tilde{S}_2\ket{\psi} +
\sum_\chi
\bra{\psi'}\tilde{S}_2\ket{\chi}\bra{\chi}\tilde{Q}_2\ket{\psi}\ ,
\end{equation}
and we specify below which states $\ket{\chi}$ appear. The transitions in the expression above are a short-hand for the regulated three-point functions, appearing at first order in perturbation theory. For example,
\begin{align}
\bra{\chi}\tilde{S}_2\ket{\psi} &= g\,\cint{C(0)}\!d\bar{x}\,\int d^2y\,\llangle \chi^\dagger(\infty)\,\tilde{G}'^+(\bar{x})\,\Phi(y,\bar{y})\,\psi(0)\rrangle \nonumber\\
&= \pi g\,\llangle \chi^\dagger(\infty)\,\Phi_{\mrm{L}}(1)\,\psi(0)\rrangle\ ,
\end{align}
where $\Phi_{\mrm{L}}=G^-_{-\frac{1}{2}}\sigma_2$.  Similarly,
\begin{equation}
\bra{\chi}\tilde{Q}_2\ket{\psi} = \pi g\,\llangle \chi^\dagger(\infty)\,\Phi^\dagger_{\mrm{L}}(1)\,\psi(0)\rrangle\ .
\end{equation}
We note that these correlators do not need to be integrated, and each of them can be calculated as explained in \cite{Gaberdiel:2023lco}. In particular, we can lift the correlator to the covering surface where the fields are free and the
calculation reduces to Wick contractions;\footnote{For the fermions, the spin field complicates things slightly.} the relevant covering maps are
\begin{equation}
\bm{\Gamma}_+(z) = (w+1)\,z^w-w\,z^{w+1}\ ,\qquad \bm{\Gamma}_-(z) =
\frac{z^w}{w\,z-(w-1)}\ ,
\end{equation}
for transitions into the $w+1$ and $w-1$-twisted sector, respectively.

The main difficulty of this method of calculation comes from the enumeration of the possible states and the calculation of a large number of correlation functions. We have automated this process with Mathematica.

\subsection{Intermediate states}\label{sec:intermediate}

Let us explain how to identify the relevant intermediate states with the example of $\bra{\chi}\tilde{S}_2\ket{\psi}$, where $\ket{\psi}$ is one of the reference $2$-magnon states in (\ref{2magnon}). We start by noting that
$\ket{\chi}$ must again be a $\frac{1}{4}$-BPS state \cite{Keller:2019suk,Benjamin:2021zkn}. In addition, its dimension must be the same as that of $\ket{\psi}$, and we have the restrictions that come from the fact that the perturbation preserves various charges, see Appendix \ref{app:conserved_charges} for more details. Given that the operator $\tilde{S}_2$ has right-moving R-charge and dimension $1/2$, it maps into the $w+1$-twisted sector.\footnote{$\tilde{S}_2$ cannot map into the $(w-1)$-twisted sector due to our choice of considering states whose right-moving part is the top BPS state in the $w$-cycle twisted sector. Note that there is a `multi-trace' BPS state with the correct charges in the $(w-1)$-twisted sector, namely the state $\ket{w-1}\otimes K^+_{-1}\ket{0}\otimes \ket{0}^{N-w}$. However, transitions to this state are $1/\sqrt{N}$ suppressed, and hence do not contribute in the planar (large $N$) limit.} Furthermore, it is positively charged under the diagonal residual $\mfr{su}(2)$ which rotates the bosons and is preserved by the perturbation (see Appendix \ref{app:conserved_charges}). This means that under $\tilde{S}_2$ we need to look at transitions to the states of the schematic form
\begin{equation}
\begin{array}{l lc}
\alpha^2\psi^-\ket{w+1}\ ,\quad & \\
\alpha^2\alpha^2\bar{\alpha}^1\psi^-\ket{w+1}\ ,\quad &  \alpha^1\alpha^2\bar{\alpha}^2\psi^-\ket{w+1}\ ,\\
\alpha^2\bar{\psi}^+\psi^-\psi^-\ket{w+1}\ ,\quad &  \alpha^2\psi^+\bar{\psi}^-\psi^-\ket{w+1}\ ,\\
\alpha^1\alpha^2\alpha^2\bar{\psi}^-\ket{w+1}\ ,\quad & \alpha^2\psi^+\psi^-\psi^-\ket{w+1}\ ,\\
\qquad \dots &  \qquad\dots
\end{array}
\end{equation}
where we have listed all the states with four or less magnon excitations. Only the first line describes the states considered in the original description of \cite{Gaberdiel:2023lco}, but in general, transitions to all of these states are non-zero. The effect of these transitions on the anomalous dimension matrix $\tilde{\gamma}$ is strongest on the diagonal, as they constructively interfere and shift the values upwards. The off-diagonal elements of $\tilde{\gamma}$ are still $1/w$ suppressed.

In particular, this is the case for the matrix elements that mix one of the original $2$-magnon states to a state $\ket{\chi}$ that involves $k\geq 4$ magnons. As in \cite{Gaberdiel:2023lco}, see also the ancillary notebook, the corresponding matrix element can be obtained by performing the different field contractions. Since we work to first order in the perturbation, all but one contraction will be of `same-species' type, coming from a Wick contraction of a pair of conjugate fields on the covering surface. For the calculation of the matrix element between a state with two magnons and a state $\ket{\chi}$ with $k\geq 4$ magnons, we therefore necessarily have at least $k/2-1$ same-species contractions between magnons of $\ket{\chi}$. These `self-contractions', say between magnons with mode numbers $p=m/w$ and $q=n/w$, are proportional to
%
\begin{equation}\label{eq:magnon_number_violation_suppression}
\frac{1}{w}\,\frac{\sin(\pi p)}{p+q}\ .
\end{equation}
As in the steps leading to eq.~(5.3) in \cite{Gaberdiel:2023lco}, the denominator $p+q$ can only become small if $p,q$ both go to zero, but this is cancelled by the $\sin(\pi p)$ in the numerator. As a consequence, this term is proportional to $\frac{1}{w}$, and the whole matrix element therefore scales as $w^{1-k/2}$.

In addition to including these higher magnon states, one also needs to take into account that $\tilde{Q}_2$ cannot only map into the $(w-1)$-twisted sector, but also into the $(w+1)$-twisted sector. This follows because we work with the top BPS state $\ket{w}$, and in the $(w+1)$-twisted sector there is a BPS state $\ket{w+1}_-$ with charge $\frac{w}{2}$, allowing the $\tilde{Q}_2$ transition. However, the corresponding right-moving three-point function is actually $\frac{1}{w}$ suppressed with respect to the other transitions, see Appendix \ref{app:right_moving_normalisation}.

\subsection{Consistency checks}\label{sec:cons}

Once we take all possible intermediate states into account, we find that the three-point function calculation reproduces indeed exactly the four-point function calculation, as predicted by eq.~(\ref{fullmixing}). This is a highly non-trivial consistency check on both calculations.

We have carried out further consistency checks coming from the conservation of the algebra. For example we have confirmed that $(\tilde{Q}_2)^2=0$; this is not trivially satisfied, but instead requires all possible intermediate states, once one includes some transitions beyond the $2$-magnon sector. Another important identity is $\{\tilde{S}_1,\tilde{Q}_1\}=\{\tilde{S}_2,\tilde{Q}_2\}$, which is also nontrivially satisfied and (indirectly) relates transitions to the $(w+1)$-twisted sector to those to the $(w-1)$-twisted sector.

Finally, we have checked that $[K^+_{-1},\tilde{Q}_2]=[K^-_1,\tilde{Q}_2]=0$, and similarly for $\tilde{S}_2$. This last check is a bit more involved, as one needs to take multi-trace states into account, as the global symmetries $K^\pm_{\mp 1}$ mix multi- and single-trace states. This can be seen as follows. The global mode $K^+_{-1}$ is the symmetric combination
\begin{equation}
K^+_{-1} = \sum_{i=1}^N K^{+\,(i)}_{-1}\ ,
\end{equation}
where $K^{+\,(i)}_{-1}$ is the mode in the $i$-th copy of the seed theory, and one has to sum over all the $N$ copies to respect the orbifold symmetry. If we apply this operator to the single-trace state $\ket{\phi}\otimes \ket{0}^{N-w}$ in the $w$-twisted sector, where we have explicitly written out the remaining $N-w$ copies that are in the ground state, one obtains the multi-trace state
\begin{equation}\bbr{K^+_{-1}\ket{\phi}}\otimes \ket{0}^{N-w}+\sum_{i=0}^{N-w}\ket{\phi}\otimes\ket{0}^i\otimes\bbr{K^+_{-1}\ket{0}}\otimes\ket{0}^{N-w-i-1}\ .
\end{equation}
Conversely, if $K^-_{1}\ket{\phi}=0$, applying $K^-_1$ to the multi-trace state $K^+_{-1}\ket{\phi}\otimes\ket{0}^{N-w}$ just gives back the single trace state $\ket{\phi}\otimes \ket{0}^{N-w}$. For the calculation of the anomalous dimensions of single-trace states, on the other hand, one does not need to take these multi-trace states into account since their contribution is proportional to $1/\sqrt{N}$ \cite{Pakman:2009zz},  and hence vanishes in the large $N$ limit.

\subsection{Algebraic structure of the eigenstates}\label{sec:3.3}

Given that the calculation of \cite{Gaberdiel:2023lco} reproduces the correct anomalous conformal dimensions, one may  expect that it also captures correctly the action of the supercharges, and this indeed turns out to be the case. In order to explain this, let us denote by $\ket{\epsilon_m}$, $m=1,\dots,\lceil \frac{w-1}{2}\rceil$, the eigenstate associated to the two-magnon state $\ket{\Psi_m}$ in eq.~(\ref{2magnon}), i.e.\ the eigenstate that has the largest overlap with $\ket{\Psi_m}$. First, we find experimentally that each $\ket{\epsilon_m}$ is annihilated by $\tilde{Q}_2$ as well as $\tilde{S}_1$, i.e.~it sits in a short representation with respect to the superalgebra. To see this, one needs to take the mixing with the $(w+2)$-twisted sector into account,\footnote{This additional complication is absent if one instead works with bosonic descendants of the \emph{bottom} BPS state on the left, where one sees this property already when only looking at mixing in the $w$-twisted sector.} which is further discussed in Appendix \ref{app:wp2_covering_map}. It turns out that this does not affect the spectrum associated to the $\ket{\epsilon_m}$, as one can check using the ancillary notebook. Note that the inclusion of this sector only modifies transitions which are subleading in $1/w$ and are neglected in the limit prescription of \cite{Gaberdiel:2023lco}. In particular, for $\tilde{Q}_2$, this only affects the $1/w$ suppressed transitions from the $w$ to the $w+1$ twisted sector.

As a consequence of the fact that $\ket{\epsilon_m}$ sits in a short representation, its anomalous dimension is simply given by
\begin{equation}\label{3.5}
\epsilon_m = \bra{\epsilon_m}\tilde{\gamma}\ket{\epsilon_m} = \big\|\tilde{S}_2\ket{\epsilon_m}\big\|^2 = \big\|\tilde{Q}_1\ket{\epsilon_m}\big\|^2\ .
\end{equation}
This is the basic algebraic structure also found in \cite{Gaberdiel:2023lco} when applying the analysis to the states $\ket{\Psi_m}$. Together with the matching of the anomalous dimensions shown in Fig.~\ref{fig:w16_spectrum}, this suggests that the full calculation is quantitatively controlled by the analysis of \cite{Gaberdiel:2023lco}.

To find more evidence for this statement we explicitly study the action of $\tilde{S}_2$ on the eigenstate $\ket{\epsilon_m}$ and compare it to the calculation in \cite{Gaberdiel:2023lco}. Let us write the (normalised) eigenstate as
\begin{equation}\label{eigen}
\ket{\epsilon_m} = \frac{1}{\mcl{N}_{w,m}}\,\Big(\tfrac{1}{\sqrt{m(w-m)}}\,\alpha^2_{-\frac{m}{w}}\alpha^2_{-1+\frac{m}{w}}\ket{w}+\text{tail of corrections}\Big)\ ,
\end{equation}
where $\mcl{N}_{w,m}$ depends on both $w$ and $m$, and is fixed by requiring the eigenstate $\ket{\epsilon_m}$ to be normalised.  The `tail' contains the other two-magnon states (with small coefficients), as well as higher magnon corrections. The operator $\tilde{S}_2$ maps this state into a linear combination of the  form
\begin{align}
\tilde{S}_2\ket{\epsilon_m} = \sum_{k=1}^{w} \frac{c(k;m)}{\sqrt{(w+1)\,k}}\, \alpha^2_{-\frac{k}{w+1}} \psi^-_{-\frac{1}{2}+\frac{k}{w+1}}\ket{w+1} + \text{tail}\ .
\end{align}
The dominant transitions are to the states with $k=m$ and $k=w+1-m$, for which the individual mode number of each magnon is approximately conserved.

Let us compare this to the calculation of \cite{Gaberdiel:2023lco}, where at finite $w$ we only considered transitions from the schematic states of the form $\alpha^2\alpha^2|w\rangle$ to those of the form $\alpha^2\psi^-|w\rangle$. We can diagonalise the anomalous dimension matrix on this subspace, and obtain the naive `eigenstates' $\ket{\epsilon^\mrm{old}_m}$ which now only contain two-magnon states,
\begin{equation}
\ket{\epsilon^\mrm{old}_m} = \frac{1}{\mcl{N}^\mrm{old}_{w,m}}\,\Big(\tfrac{1}{\sqrt{m(w-m)}}\,\alpha^2_{-\frac{m}{w}}\alpha^2_{-1+\frac{m}{w}}\ket{w}+\text{other two-magnon states}\Big)\ .
\end{equation}
(Here $\mcl{N}^{\mrm{old}}_{m}$ also depends on $m$ and $w$, and is again fixed by requiring $\ket{\epsilon^\mrm{old}_m}$ to be normalised.) In this approach, the action of $\tilde{S}_2$ takes the form
\begin{equation}
\tilde{S}^\mrm{old}_2\ket{\epsilon^\mrm{old}_m} = \sum_{k=1}^{w} \frac{c^\mrm{old}(k;m)}{\sqrt{(w+1)\,k}}\, \alpha^2_{-\frac{k}{w+1}} \psi^-_{-\frac{1}{2}+\frac{k}{w+1}}\ket{w+1} \ ,
\end{equation}
where $c^\mrm{old}(k;m)$ are the finite $w$ expressions from \cite{Gaberdiel:2023lco}, and there is obviously \emph{no tail} of multi-magnon states. We can now compare the coefficients $c(k;m)$ to $c^\mrm{old}(k;m)$, and we find that there is good quantitative agreement, already at small $w$
\begin{equation}\label{eq:coefficient_relation}
\mcl{N}_{w,m}\,c(k;m)\approx \mcl{N}^\mrm{old}_{w,m}\,c^\mrm{old}(k;m)\ .
\end{equation}
Based on the agreement of the spectra in Section \ref{sec:full_AD}, we expect the agreement to become better for large $w$. However, the largest value of $w$ for we can analyse this problem for the states of the form (\ref{2magnon}) is $w=6$. (For that case, the agreement between the coefficients is 1-3 \%, at least for the two values of $k$ for which either $\mcl{N}_m\,c(k;m)$ or $\mcl{N}^\mrm{old}_m\,c^\mrm{old}(k;m)$ is bigger than $0.1$.)

\begin{figure}[h]
\begin{minipage}[t]{.49\textwidth}
\includegraphics[width=\textwidth]{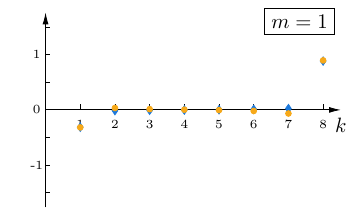}
\end{minipage}
\hfill
\begin{minipage}[t]{.49\textwidth}
\includegraphics[width=\textwidth]{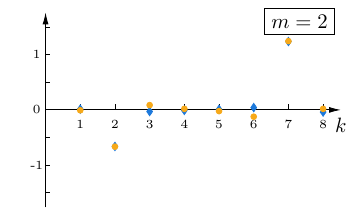}
\end{minipage}
\begin{minipage}[t]{.49\textwidth}
\includegraphics[width=\textwidth]{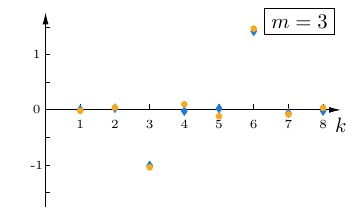}
\end{minipage}
\hfill
\begin{minipage}[t]{.49\textwidth}
\includegraphics[width=\textwidth]{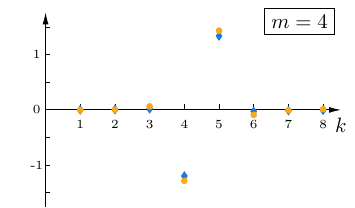}
\end{minipage}
\caption{The coefficients $\mcl{N}_{8,m}^\text{old}\,c^\text{old}(k;m)$ (blue diamonds) and $\mcl{N}_{8,m}\,c(k;m)$ (orange circles) for the eigenstates associated to $\alpha^2_{-m/8}\alpha^2_{-1+m/8}\ket{8}_-$ and $m=1,2,3,4$ from top left to bottom right.}\label{fig:coefficient_comparison}
\end{figure}

However, we can effectively push the analysis a little bit higher if instead of (\ref{2magnon}) we consider the states
\begin{equation}\label{2magnonp}
|\Psi_m\rangle \equiv \tfrac{1}{\sqrt{m(w-m)}}\alpha^2_{-\frac{m}{w}}\alpha^2_{-1+\frac{m}{w}}\ket{w}_-\ ,
\end{equation}
where $\ket{w}_-$ is the BPS state with charge $\frac{w-1}{2}$ in the $w$-twisted sector.\footnote{As far as the analysis of \cite{Gaberdiel:2023lco} is concerned, the large $w$ algebra and the anomalous dimension of those states is exactly the same as for those based on $|w\rangle$.} Then, because of charge conservation, these states can only mix with states in the $(w-2)$-twisted sector, and this allows us to push the analysis to $w=8$, see Figure \ref{fig:coefficient_comparison}.
%
%
%
Together with the fact that, see Fig.~\ref{fig:w16_spectrum},
\begin{equation}
C\,\epsilon_m \approx \epsilon^\mrm{old}_m \ ,
\end{equation}
where $C\approx 0.41$ is some $m$-independent constant of proportionality (which can be absorbed in a rescaling of the coupling), and using the expression for the anomalous dimension as the norm of $\tilde{S}_2\ket{\epsilon_m}$ squared, see eqs.~(\ref{3.5}) and (\ref{eq:coefficient_relation}), one finds
\begin{equation}
C\,\left(\frac{\mcl{N}^\mrm{old}_{w,m}}{\mcl{N}_{w,m}}\right)^2\sum_k |c^\mrm{old}(k;m)|^2 + C\,\|\text{tail}_m\|^2 \approx \sum_k |c^\mrm{old}(k;m)|^2\ .
\end{equation}
Thus it follows that the complicated `tail' only renormalises the `head' of the state. This is the  more microscopic reason why the limit of Section~\ref{sec:old}, see  \cite{Gaberdiel:2023lco}, captures the full analysis: it keeps track of the heads of the eigenstates from which one can read off the action of the supercharges, as well as the anomalous conformal dimensions. One would therefore expect that the $\frac{1}{w}$ corrections can be absorbed into a redefinition of the supercharges (and possibly also the magnon operators), so that they satisfy, in the large $w$ limit, the same relations as in \cite[eqs.~(5.24)]{Gaberdiel:2023lco}. In particular, this would then make the integrable structure of the problem manifest.

\section{Conclusions}\label{sec:concl}

In this paper we have studied in detail the perturbation in the symmetric product orbifold that is dual to turning on R-R flux in the ${\rm AdS}_3$ background. We have mainly focused on one two-magnon reference state, see eq.~(\ref{2magnon}), for which we have carried out the full calculation of the anomalous mixing matrix in the planar limit for twisted sectors up to $w=16$, see Fig.~\ref{fig:w16_spectrum}. From this we have then extracted the dispersion relation for eigenstates associated to the two-magnon states, and we have shown that the spectrum reproduces (for large $w$) the one predicted in \cite{Gaberdiel:2023lco}. (In turn, the results of \cite{Gaberdiel:2023lco} reproduced the integrable predictions of \cite{Babichenko:2009dk,Hoare:2013lja,Borsato:2013qpa,Lloyd:2014bsa,Frolov:2023pjw}.)
Our analysis is therefore an independent consistency check of the results of \cite{Gaberdiel:2023lco}, and in particular, justifies the large $w$ limit that was taken there. It also suggests that the limit of \cite{Gaberdiel:2023lco} is naturally embedded in the full calculation, see the discussion in Section~\ref{sec:3.3}, and it would be interesting to explore this more systematically.

We have carried out the calculation of the anomalous dimensions in two ways, once by contour-integrating a four-point correlator, see Section~\ref{sec:4point}, and once by summing over all possible three-point functions (OPE expansion), see Section~\ref{sec:3point}. These calculations agree perfectly. Both methods can be applied to calculate the anomalous dimensions of any quarter BPS states in the planar limit of the orbifold, and may therefore also be useful in other contexts. (The relevant Mathematica notebooks are included as ancillary files in the {\tt arXiv} submission.)

\section*{Acknowledgements} This paper is partially based on the Master thesis of one of us (F.L.). We thank Rajesh Gopakumar for initial collaboration and comments on a draft version of this paper, and Frank Coronado, Artyom Lisitsyn, Kiarash Naderi, and Vit Sriprachyakul for useful conversations.  The work of BN is supported through a personal grant of MRG from the Swiss National Science Foundation. The work of the group at ETH is also supported in part by the Simons Foundation grant 994306 (Simons Collaboration on Confinement and QCD Strings), as well the NCCR SwissMAP that is also funded by the Swiss National Science Foundation.

\appendix

\section{Conventions}\label{app:conventions}

We denote the left-moving four bosons and four fermions of the $\mbb{T}^4$ by $\alpha^i,\bar{\alpha}^i$, $i=1,2$ and $\psi^\pm,\bar{\psi}^\pm$, respectively. They satisfy the OPE relations
\begin{equation}
    \bar{\alpha}^i(x)\alpha^j(y)\sim \frac{\epsilon^{ij}}{\big(x-y\big)^2}\ ,\qquad \bar{\psi}^\pm(x)\psi^\mp(y)\sim \frac{\pm 1}{x-y} \ .
\end{equation}
Right-movers are always denoted by a tilde. The (left-moving) $\mcl{N}=4$ generators are built out of these fields as
\begin{equation}\label{N4fields}
\begin{array}{cclccl}
G^+ &= & \bar{\alpha}^2\,\psi^++\alpha^2\,\bar{\psi}^+\ , \qquad & K^+ &= & \bar{\psi}^+\,\psi^+\ , \\
G'^+ &= & -\bar{\alpha}^1\,\psi^+-\alpha^1\,\bar{\psi}^+\ ,\qquad & K^- &= & -\bar{\psi}^-\,\psi^-\ , \\
G^- &= & \bar{\alpha}^1\,\psi^-+\alpha^1\,\bar{\psi}^-\ , \qquad & K^3 &= & \frac{1}{2}:\bar{\psi}^+\,\psi^-+\bar{\psi}^-\,\psi^+:\ , \\
G'^- &= & \bar{\alpha}^2\,\psi^-+\alpha^2\,\bar{\psi}^-\ , \qquad & &&
\end{array}
\end{equation}
and
\begin{equation}
L =  \,:\bar{\alpha}^1\,\alpha^2-\bar{\alpha}^2\,\alpha^1:+\,\frac{1}{2}:\psi^+\,\partial \bar{\psi}^- + \bar{\psi}^-\,\partial \psi^+ - \bar{\psi}^+\,\partial\psi^- - \psi^-\,\partial\bar{\psi}^+:\ .
\end{equation}
These fields generate the $c=6$ ${\cal N}=4$ superconformal algebra, whose modes satisfy
\begin{equation}
\begin{array}{rcl}
{}   [L_m,L_n]&=& (m-n)L_{m+n}+\tfrac{1}{2}(m^3-m)\delta_{m+n,0}\ ,\\
{}   [L_m,K^a_n]&= & -nK^a_{m+n}\ ,\\
{}   [L_m,G^a_r]&= & \big(\tfrac{m}{2}-r\big)G^a_{m+r}\ ,\\
{}   [L_m,G'^a_r]&= & \big(\tfrac{m}{2}-r\big)G'^a_{m+r}\ ,\\
{}   \{G^+_r,G^-_s\}&= & \{G'^+_r,G'^-_s\}=L_{r+s}+(r-s)K^3_{r+s}+\big(r^2-\tfrac{1}{4}\big)\delta_{r+s,0}\ ,\\
{}   \{G^\pm_r,G'^\pm_s\}&= & \mp(r-s)K^\pm_{r+s}\ ,
\end{array}
\end{equation}
as well as
\begin{equation}
\begin{array}{cclccl}
{}[K^3_m,K^3_n] &=& \tfrac{m}{2}\delta_{m+n,0}\ , \qquad   & [K^3_m,G^\pm_r]&=& \pm\tfrac{1}{2}G^\pm_{m+r}\ , \\
{} [K^3_m,K^\pm_n]&=& \pm K^\pm_{m+n}\ ,\qquad & [K^3_m,G'^\pm_r]&= & \pm\tfrac{1}{2}G'^\pm_{m+r}\ , \\
{}[K^+_m,K^-_n]&= & 2K^3_{m+n}+m\,\delta_{m+n,0}\ ,\qquad & & &
\end{array}
\end{equation}
and
\begin{equation}
\begin{array}{cclccl}
{}[K^+_m,G^-_r] &=& G'^+_{m+r}\ , \qquad & [K^+_m,G'^-_r] &=& -G^+_{m+r}\ ,\\
{}[K^-_m,G^+_r] &=& -G'^-_{m+r}\ , \qquad & [K^-_m,G'^+_r] &=& G^-_{m+r}\ .
\end{array}
\end{equation}
The other (anti-)commutators vanish.

\subsection{Conserved charges}\label{app:conserved_charges}
The perturbation field is
\begin{equation}
\Phi = \frac{i}{\sqrt{2}}\,\left(G^-_{-\frac{1}{2}}\tilde{G}'^-_{-\frac{1}{2}} -G'^-_{-\frac{1}{2}}\tilde{G}^-_{-\frac{1}{2}} \right)\sigma_2\ ,
\end{equation}
where $\sigma_2$ is the $h=\frac{1}{2}$ BPS state in the 2-cycle twisted sector (with spin up). This perturbation preserves the following charges:
\begin{itemize}
\item The number of barred magnons. The bar charge acts as
\begin{equation}
[U_{\mrm{bar}},A]=A\ ,\quad [U_{\mrm{bar}},\bar{A}]=-\bar{A}\ ,
\end{equation}
where $A$ stands for any free boson or fermion. This charge is conserved because all of the supercharges are neutral under $U_{\mrm{bar}}$,
\begin{equation}
[U_{\mrm{bar}},G^\pm_r] = 0\ \Longrightarrow \ [U_{\mrm{bar}},\Phi]=0\ .
\end{equation}
\item The $\mfr{su}(2)$ R-charge, i.e.~the eigenvalue of $K^3_0$, under which the fermions are charged
\begin{equation}
[K^3_0,\psi^\pm]=\pm\frac{1}{2}\psi^\pm\ ,\quad [K^3_0,\bar{\psi}^\pm]=\pm\frac{1}{2}\bar{\psi}^\pm\ .
\end{equation}
\item There is a diagonal residual $\mfr{su}(2)$ charge acting on the free bosons (and leaving the fermions invariant), which is also preserved by the perturbation. The residual left-moving charge is
\begin{equation}\label{eq:res_charge}
[U_{\mrm{res}},\alpha^1] = \alpha^1\ ,\quad [U_{\mrm{res}},\alpha^2] = -\alpha^2
\ ,\quad [U_{\mrm{res}},\bar{\alpha}^1] = \bar{\alpha}^1\ ,\quad [U_{\mrm{res}},\bar{\alpha}^2] = -\bar{\alpha}^2\ .
\end{equation}
Then, the diagonal action, i.e.\ the simultaneous action on left- and right-movers, commutes with the perturbation
\begin{equation}
[U_{\mrm{res}}+\tilde{U}_{\mrm{res}},\Phi]=0\ .
\end{equation}
In practice, since we look at quarter BPS states, this means that $U_{\mrm{res}}$ is preserved in the four-point function calculation. When calculating the anomalous dimension via three-point functions, we note that $\tilde{S}_2=G'^+_{-1/2 }$ is positively charged under this symmetry, while $\tilde{Q}_2=G'^-_{1/2}$ is negatively charged. This restricts the possible intermediate states.
\end{itemize}

\section{Some calculational details}

In this section we give some more details regarding the actual computation of the four- and three-point functions.

\subsection{A $2$-magnon matrix element}\label{sec:explicit_correlator}

Here we show the basic method of the four-point calculation using covering maps.
Let us consider the matrix element $\bra{\Psi_{m_2}}\tilde{\gamma}\ket{\Psi_{m_1}}$, where as before, see eq.~(\ref{2magnon})
\begin{equation}
\ket{\Psi_{m_i}}=\frac{1}{\sqrt{m_i(w-m_i)}}\,\alpha^2_{-\frac{m_i}{w}}\alpha^2_{-1+\frac{m_i}{w}}\ket{w}
\end{equation}
are the normalised archetypal states. As discussed in Section \ref{sec:covering_surface_calculation}, these matrix elements can be calculated on the covering surface, where they are contour integrals of the form
\begin{align}
\bra{\Psi_{m_2}}\tilde{\gamma}\ket{\Psi_{m_1}} &= 2\pi^2g^2\,\cint{}dl\,\frac{dx}{dl}\,\cint{C(\infty)}\!d\zeta_1\,d\zeta_2\,\bm{\Gamma}^{\frac{m_2}{w}}(\zeta_1)\,\bm{\Gamma}^{1-\frac{m_2}{w}}(\zeta_2) \nonumber \\
&\hspace{3cm}\times \cint{C(0)}\!dz_1\,dz_2\,\bm{\Gamma}^{-\frac{m_1}{w}}(z_1)\,\bm{\Gamma}^{-1+\frac{m_1}{w}}(z_2)\,\mcl{C}\big(z_i,\zeta_i,l,\bar{l}\big)\ ,
\end{align}
with
\begin{align}
\mcl{C}\big(z_i,\zeta_i,l,\bar{l}\big) &= \mcl{N}(l,\bar{l})\,\Big\langle\hat{\phi}_w(\infty)\,\hat{\bar{\alpha}}^1(\zeta_1) \hat{\bar{\alpha}}^1(\zeta_2)\nonumber\\
&\hspace{2cm}\times\big(\hat{\Phi}^\dagger_{\mrm{L}}(1)\,\hat{\Phi}_\mrm{L}(u,\bar{u})
+
\hat{\Phi}'^\dagger_\mrm{L}(1)\,\hat{\Phi}'_{\mrm{L}}(u,\bar{u})
\big)\, \hat{\alpha}^2(z_1)\hat{\alpha}^2(z_2)\,\hat{\phi}_w(0)\Big\rangle\ .
\end{align}
The factor $\mcl{N}(l,\bar{l})$ will be given below, see eq.~(\ref{B.7}); it contains contributions coming from orbifold averaging, normalisation, and the right-moving part of the correlator. The correlator $\mcl{C}$ is a product of a bosonic and a fermionic correlator. The bosonic contribution is given by Wick contractions. Using the definition of the $\Phi_{\mrm{L}}$ and the conventions in Appendix \ref{app:conventions}, one finds
\begin{align}
\mcl{C}_{\mrm{bos}}\big(z_i,\zeta_i,l\big) &= \frac{4}{(1-u)^2}\,\left(\frac{1}{(\zeta_1-z_1)^2}\frac{1}{(\zeta_2-z_2)^2} +(z_1\leftrightarrow z_2)\right)\nonumber\\
&\quad + \left(\frac{1}{(\zeta_1-1)^2}\frac{1}{(u-z_1)^2}\frac{1}{(\zeta_2-z_2)^2}+\text{perms.}\right)\nonumber\\
&\quad + \left(\frac{1}{(\zeta_1-u)^2}\frac{1}{(1-z_1)^2}\frac{1}{(\zeta_2-z_2)^2}+\text{perms.}\right)\ .
\end{align}
The fermionic part of the correlator can, for example, be calculated using a bosonisation of the fermions, see  Appendix \ref{app:bosonisation}. For the correlator at hand, it is independent of $\zeta_i$ and $z_i$. The matrix element is thus of the form
\be
\bra{\Psi_{m_2}}\tilde{\gamma}\ket{\Psi_{m_1}} = 2\pi^2g^2\,\cint{}dl\,\frac{dx}{dl}\,\mcl{N}(l,\bar{l})\,\mcl{C}_{\mrm{ferm}}(l)\, F(l)\ ,
\ee
where
\be
F(l) = \cint{C(\infty)}\!d\zeta_1\,d\zeta_2\,\bm{\Gamma}^{\frac{m_2}{w}}(\zeta_1)\,\bm{\Gamma}^{1-\frac{m_2}{w}}(\zeta_2)\, \cint{C(0)}\!dz_1\,dz_2\,\bm{\Gamma}^{-\frac{m_1}{w}}(z_1)\,\bm{\Gamma}^{-1+\frac{m_1}{w}}(z_2)\,\mcl{C}_{\mrm{bos}}\big(z_i,\zeta_i,l,\bar{l}\big) \ .
\ee
The factor in front of $F(l)$ has a simple $l$-dependence,
\begin{align}\label{B.7}
\frac{dx}{dl}\,\mcl{N}(l,\bar{l})\,\mcl{C}_{\mrm{ferm}}(l) &= \frac{1}{\sqrt{m_1 m_2 (w-m_1) (w-m_2)}}\times \frac{l}{w\big(w\,l-(w-1)\big)} \nonumber\\
&\quad\times\frac{w-1-2w\bar{l}}{2 w (w+1)
(\bar{l}-1)
\big((w+1)\bar{l}-(w-1)\big)}\ ,
\end{align}
where the first factor in the first line is due to the normalisation of the states, and the second line is the right-moving contribution to the correlator.

The details of the function $F(l)$ are more complicated, but its structure is straight-forward. As $\mcl{C}_{\mrm{bos}}$ is given by Wick contractions, the integrated correlator is made up of building blocks where one contour-integrates functions of the form $\frac{1}{(z-\rho)^2}$ with factors of $\bm{\Gamma}(z)$ to some fractional power. These can be expressed explicitly in terms of (iterated) sums of rational functions in $l$. An easy example is the case when an $\alpha^2$ inserted at $0$ contracts with the perturbation at $1$,
\begin{align}
\cint{C(0)}\!dz\,\bm{\Gamma}^{-\frac{m}{w}}(z)\,\frac{1}{(1-z)^2} &= \left(\frac{(w+1)\,l-w}{l}\right)^{-\frac{m}{w}}\,\sum_{k=0}^{m-1} b_{m,k}(l)\times(m-k)\ ,\nonumber\\
b_{m,k}(l) &= \sum_{r=0}^k \binom{-\frac{m}{w}}{r}\binom{\frac{m}{w}}{k-r}\,\left(\frac{w\,l-(w-1)}{(w+1)\,l - w}\right)^r\,l^{k-r}\ .
\end{align}
This can be found by Taylor expanding $\bm{\Gamma}^{-\frac{m}{w}}(z)$ around $z=0$ (where the function is single valued). The other building blocks can be expressed similarly.

\subsection{The \texorpdfstring{$(w+2,2,2,w)$}{(w+2,2,2,w)} covering map}\label{app:wp2_covering_map}
The states we consider in the $w$-twisted sector are quarter BPS states of the form
\begin{equation}
\ket{\phi}\otimes \ket{\widetilde{w}} \ .
\end{equation}
Here $\ket{\phi}$ stands for an arbitrary left-moving state, and we have explicitly written down the right-moving component $\ket{\widetilde{w}}$, which is the top BPS state with charge and dimension $\frac{w+1}{2}$. The perturbation mixes all states with the same charges; besides the states in the $w$-twisted sector, there are also some in the $w+2$-twisted sector of the form
\begin{equation}
\ket{\phi'}\otimes\big(\tilde{K}^-_{1}\ket{\widetilde{w+2}}\big)\ .
\end{equation}
Mixing between the different twisted sectors is small and turns out to only minimally affect the spectrum. However, for completeness we also consider this contribution. For the four-point function calculation, we thus also need the covering map for the correlators with ramification profile $(w+2,2,2,w)$, with twist operators sitting at $\infty, x, 1, 0$. It is given by

\begin{equation}
\bm{\Gamma}_{++}(z) = \frac{w\,-(w+1)}{(b-1)^2}\, z^w (z-b) \left(z-\frac{(w+1)\,b -(w+2)}{w\,b -(w+1)}\right)\ .
\end{equation}
The ramification points of this map are $\infty, v(b),1,0$, where
\begin{equation}
v(b) = \frac{w\,b\, \big((w+1)\,b-(w+2)\big)}{(w+2) \big(w\,b - (w+1)\big)}\ ,
\end{equation}
which corresponds to an insertion of the perturbation at $x = \bm{\Gamma}_{++}\big(v(b)\big)$ in the base space. Integration over the position of the perturbation in the base space then corresponds to integrating over $b$.\footnote{There is an automorphism $b\mapsto \frac{(w+1)b-(w+2)}{w b - (w+1)}$ which leaves the covering map invariant. This needs to be taken into account with appropriate symmetry factors for the residues.}

\subsection{Twist-dependent normalisation factors}\label{app:right_moving_normalisation}

To obtain the correct weighting of three-point functions, one needs to take the right-moving BPS correlators into account. The relevant normalisation factors are
\begin{align}
\bra{w+1}\sigma_2(1)\ket{w} &= \sqrt{\frac{w}{w+1}}\ ,\nonumber\\
{}_-\!\!\bra{w+1}\sigma^\dagger_2(1)\ket{w} &= \frac{1}{\sqrt{w(w+1)}}\ ,\nonumber\\
\bra{w-1}\sigma^\dagger_2(1)\ket{w} &= \sqrt{\frac{w-1}{w}}\ ,\\
{}_-\!\!\bra{w+1}\sigma_2(1)\ket{w}_- &= \sqrt{\frac{w+1}{w}}\ .\nonumber
\end{align}
All other relevant correlators can be obtained from these by conjugation.

There are also similar factors for the left-moving part of the correlator. Taking also the orbifold averaging factors, see \cite{Pakman:2009zz}, into account, one obtains a factor $\sqrt{w/(w+1)}$ for transitions into the $w+1$ twisted sector, and $\sqrt{(w-1)/w}$ for transitions into the $w-1$ twisted sector.

\subsection{Bosonisation and cocycle factors}\label{app:bosonisation}

When calculating the correlation functions on the covering surface, spin fields appear due to the twist-2 perturbation. One way to deal with this is by bosonising the fermions on the covering surface. Especially when matching between the four- and three-point function, it turns out that one needs to be careful about keeping track of the cocycle factors. We implement them in an algebraic way as follows.

We bosonise the fermions as
\begin{equation}
\psi^+ = c_{1,0}\,e^{i\varphi}\ ,\quad \bar{\psi}^- = -c_{-1,0}\,e^{-i\varphi}\ ,\quad \bar{\psi}^+ = c_{0,1}\,e^{i\varphi'}\ ,\quad \psi^-=c_{0,-1}\,e^{-i\varphi'}\ ,
\end{equation}
where $\varphi(z)\varphi(\zeta)=-\log(z-\zeta)$, $\varphi'(z)\varphi'(\zeta)=-\log(z-\zeta)$ and the $c_{\epsilon_1,\epsilon_2}$ are cocycle factors. These factors satisfy the simple relations
\begin{equation}
c_{\pm 1,0}\,c_{\mp 1,0} = c_{0,\pm 1}\,c_{0,\mp 1} = 1 \ ,\qquad c_{\pm 1,0} \, c_{0,\pm 1} = - c_{0,\pm 1}\, c_{\pm 1,0}\ .
\end{equation}
The $h=\frac{1}{2}$ spin up BPS state in the 2-twisted sector is lifted to
\begin{equation}
\hat{\phi}_2 = c_{\frac{1}{2},\frac{1}{2}}\,e^{\frac{i}{2}(\varphi+\varphi')}\ ,
\end{equation}
while its hermitian conjugate is
\begin{equation}
\hat{\phi}^\dagger_2 = c_{-\frac{1}{2},-\frac{1}{2}}\,e^{-\frac{i}{2}(\varphi+\varphi')}\ .
\end{equation}
These spin-field cocycles are related by
\begin{equation}
c_{\frac{1}{2},\frac{1}{2}} = c_{1,0}\,c_{0,1}\,c_{-\frac{1}{2},-\frac{1}{2}}\ .
\end{equation}
We further specify the relations
\begin{equation}
c_{\frac{1}{2},\frac{1}{2}}\,c_{-\frac{1}{2},-\frac{1}{2}} = 1\ ,\quad c_{-\frac{1}{2},-\frac{1}{2}}\,c_{1,0} = -i\,c_{1,0}\,c_{-\frac{1}{2},-\frac{1}{2}}\ ,\quad c_{-\frac{1}{2},-\frac{1}{2}}\,c_{0,1} = -i\,c_{0,1}\,c_{-\frac{1}{2},-\frac{1}{2}}\ .
\end{equation}
Together, this gives all the necessary relations to calculate the correlators on the covering surface with the correct statistics.

\end{document}